\documentclass[reprint,showpacs,amsmath,amssymb,aps]{revtex4}
\newcommand{\Array}[2]{\left(\begin{array}{#1}#2\end{array}\right)}

\begin{document}

\title{
CP violation in the CKM mixing for degenerate quark masses}
    \author{Ying Zhang\footnote{E-mail:hepzhy@mail.xjtu.edu.cn.}}
    \address{School of Science, Xi'an Jiaotong University, Xi'an, 710049, China}

\begin{abstract}
CP violation in the CKM mixing is discussed for the case of quark mass degeneracy that is approximate in the quark mass hierarchy limit.
Differing from the traditional understanding of CP vanishing for degenerate masses, we find degenerate symmetry plays a non-trivial role in CP violation. The minimal flavor structure model is reviewed to demonstrate the role of degenerate symmetry in quark flavor mixing, particularly in CP violation. This relation between mass hierarchy and CP violation helps us understand the origin of CP violation and assists the construction of the flavor model. 
\end{abstract}


\maketitle

\section{Motivation}
The standard model (the SM) has successfully described $SU(3)_c\times SU(2)_L\times U(1)_Y$ gauge interactions of the strong and electroweak interactions in a concise mathematical form with simple guage couplings. However, the grace of gauge interactions does not apply to Yukawa interactions \cite{Hocker2006,Isidori2010,Zupan2019}. Yukawa couplings in the SM are a 3-order complex matrix for each kind of quarks and charged leptons, which govern all flavor phenomenology: mass spectrum and flavor mixing.  
Due to these unclear and redundant Yukawa couplings, the relation between fermion masses and flavor mixing is still unknown.

In the quark sector, up-type/down-type quark masses have a hierarchal structure 
	\begin{eqnarray*}
	h_{12}^q\equiv \frac{m_1^q}{m_2^q}\ll 1
	,~~~~
	h_{23}^q\equiv \frac{m_2^q}{m_3^q}\ll 1,~~~~{\rm for}~~q=u,d
	\end{eqnarray*}
This is a good approximation to explore quark flavor structure.
 The CKM matrix should keep its values approximately in the hierarchy limit, providing a key clue to decoding quark mixing and CP violation. (A similar case is also discussed in the lepton sector \cite{Raidal2008}.)
Frequently, it is cursorily believed that the presence of degenerate mass is a sufficient condition for CP violation to vanish \cite{Jarlskog1985}. 
This point of view inevitably meets a challenge in explaining why the CP violating phase in CKM mixing has a large value rather than a small one as a perturbation correction from the mass hierarchy.

In the paper, we focus on the relation between mixing matrix and CP violation in the case of mass degeneracy. After briefly reviewing Jarlskog's original 1986 work in Sec. \ref{sec.degeneracy}, we highlight two problems that require attention regarding CP vanishing in mass degeneracy.
In Sec. \ref{sec.degeneratesym}, we illustrate how a degenerate $SU(2)$ symmetry contributes non-trivially to CKM mixing. Discussion on generating CP violating phase from a real mixing matrix is presented, expressing non-vanishing CP violation in case of mass degeneracy. We also review the minimal flavor structure proposed in recent research on flavor structure in Sec. \ref{sec.MFS}. The role of degenerate symmetry in the CP violating phase is discussed. A summary is provided in Sec. \ref{sec.summ}.

\section{Quark Mixing for Degeneracy}
\label{sec.degeneracy}
Considering the squared mass matrix $M_{SQ}^q\equiv M^q(M^q)^\dag$ with $q=u,d$ for up-type and down-type quarks, it can be diagonalized   by left-handed transformation $U_L^q$ as
	\begin{eqnarray}
		U_L^q M_{SQ}^q (U_L^q)^\dag={\rm diag}\Big((m_1^q)^2,(m_2^q)^2,(m_3^q)^2\Big)
		\label{eq.UMU00}
	\end{eqnarray}
Defining a commutator
	\begin{eqnarray}
		\Big[ M_{SQ}^u,M_{SQ}^d\Big]=iC
		\label{eq.defC}
	\end{eqnarray}
	the determinant of $C$ has been given in \cite{Jarlskog1985} as
	\begin{eqnarray}
		\det[C]&\!\!\!\!\!=\!\!\!\!\!&-2[(m_3^u)^2-(m_2^u)^2][(m_3^u)^2-(m_1^u)^2][(m_2^u)^2-(m_1^u)^2]
		\nonumber\\
		&&\times[(m_3^d)^2-(m_2^d)^2][(m_3^d)^2-(m_1^d)^2][(m_2^d)^2-(m_1^d)^2]J_{CP}
		\label{eq.DetC}
	\end{eqnarray}
	with Jarlskog invariant
	\begin{eqnarray}
		J_{CP}=Im[V_{11}V_{22}V_{12}^*V_{21}^*]
	\end{eqnarray}
Here, $V_{ij}$ is an element of quark CKM mixing matrix defined by $V=U_L^u(U_L^d)^\dag$.
Using the invariance of $\det[C]$, $J_{CP}$ can be expressed by the standard CKM mixing angles and CP violating phase as
	\begin{eqnarray}
		J_{CP}=s_{13}c_{13}^2s_{23}c_{23}s_{12}c_{12} s_{\delta}
		\label{eq.Jcpmixingangle}
	\end{eqnarray}
with $s_{ij}=\sin\theta_{ij},c_{ij}=\cos\theta_{ij},s_\delta=\sin\delta_{CP}$. Here, the standard CKM matrix is expressed by
	\begin{eqnarray}
		V=\Array{ccc}{1& 0 & 0 \\ 0 & c_{23} & s_{23} \\ 0 & -s_{23} & c_{23}}
		\Array{ccc}{c_{13} & 0 & s_{13}e^{-i\delta_{CP}} \\ 0 & 1 & 0 \\ -s_{13}e^{i\delta_{CP}} & 0 & c_{13}}
		\Array{ccc}{c_{12} & s_{12} & 0 \\ -s_{12} & c_{12} & 0 \\ 0 & 0 & 1}
		\label{eq.DefCKM}
	\end{eqnarray}

Using Eqs. (\ref{eq.DetC}) and (\ref{eq.Jcpmixingangle}), CP conservation, i.e. $\delta_{CP}=0$, can lead to a vanishing $\det[C]$. However, the latter is not a sufficient condition of no CP violation. An actual situation appears in mass degeneracy as an approximation of the mass hierarchy limit. 
If quark mass degeneracy leads to vanishing CP violation, then $\delta_{CP}$ must be a small quantity resulting from quark mass hierarchy correction.
In mathematical, $\delta_{CP}$ can generally be expanded in terms of hierarchy $h^q_{23}$
	\begin{eqnarray}
	\delta_{CP}= c_0 + c_{1}h^q_{23} +  O(h^2) 
	\end{eqnarray}
with cofficient $c_i$. In a traditional perspective, the coefficient $c_0$ must approach zero as quark masses become degenerate, which leads to a small value of $\delta_{CP}$ as a perturbation from mass hierarchy. 
However, the current experiment value of $\delta_{CP}$ is about $65^\circ$ \cite{PDG2022}, which is too large to be regarded as a small one coming from the hierarchy contribution.

 There is additional doubt of $\delta_{CP}$ vanishing for degenerate masses, which comes from a theoretical analysis of the CKM matrix. Quark mixing matrix $V$ is symmetrically  determined by up-type quark transformation $U_L^u$ and down-type one $U_L^d$ as
		\begin{eqnarray*}
			V=U_L^u(U_L^d)^\dag
		\end{eqnarray*}
Here, transformation $U_L^q$ transforms gauge basis to mass basis (labelled by superscript $^{(m)}$) $q_L=(U_L)^\dag q_L^{(m)}$. If assuming a mass degeneracy only in up-type quarks rather than down-type quarks, we can choose a basis on which $M^u_{SQ}$ is diagonal and the CKM mixing matrix can be expressed by $V=(U_L^d)^\dag$, which is not relative to up-type quark $U_L^u$.  The information about the degeneracy of up-type quarks is shielded, and $V$ is not affected by it. 
So, we have to seek a new way to explain the relation between the current large CP violating phase and quark mass hierarchy.

\section{Degenerate Symmetry}
\label{sec.degeneratesym}
For degenerate eigenvalues in $M_{SQ}^q$, there exists a  transformation $G^q$ that keeps the  mass eigenvalues invariant in degenerate subspace 
	\begin{eqnarray}
	G^q{\rm diag}\Big((m_1^q)^2,(m_2^q)^2,(m_3^q)^2\Big){G^f}^\dag={\rm diag}\Big((m_1^q)^2,(m_2^q)^2,(m_3^q)^2\Big)
	\label{eq.GmassG}
	\end{eqnarray}
In this section, we discuss the role of $G^q$ on the CKM mixing matrix and emphasize $\delta_{CP}$.

Using Eqs. (\ref{eq.UMU00}) and (\ref{eq.GmassG}), $M_{SQ}^q$ is generally diagonalized by transformation $G^qU_L^q$ as
		$$\Big[G^qU_L^q\Big] M_{SQ}^q\Big[G^qU_L^q\Big]^\dag={\rm diag}\Big((m_1^q)^2,(m_2^q)^2,(m_3^q)^2\Big)$$
The mixing matrix can be expressed
			\begin{eqnarray}
			V=G^uU_L^u{U_L^d}^\dag {G^d}^\dag
			\label{eq.UckmGG}
			\end{eqnarray}
The formula shows the contributions of $G^u$ and $G^d$ to quark mixing. Even in a stricter condition, i.e., the commutator $C=0$ in Eq. (\ref{eq.defC}) instead of $\det[C]=0$, the degenerate symmetry $G^q$ can still cause significant mixing of quarks. Assuming $C=0$, $M^{u}_{SQ}$ and $M^{d}_{SQ}$ have common eigenstates, labeled by column vector $v_i$ for $i=1,2,3$. We have
        \begin{eqnarray}
        \Big(v_1,v_2,v_3\Big)^\dag M^{q}_{SQ}\Big(v_1,v_2,v_3\Big)=\Big(v_1,v_2,v_3\Big)^\dag\Array{ccc}{(m_i^q)^2 && \\ & (m_i^q)^2& \\
&& (m_i^q)^2}\Big(v_1,v_2,v_3\Big)
		\end{eqnarray}
So, diagonalization transformations $U_L^y$ and $U_L^d$ are determined by
	\begin{eqnarray}
	U_L^u=U_L^d=\Big(v_1,v_2,v_3\Big)^\dag
	\end{eqnarray}
and  Eq. (\ref{eq.UckmGG}) becomes
		\begin{eqnarray}
		V=G^u (G^d)^\dag
		\label{eq.UckmGuGd}
		\end{eqnarray}
Due to the non-equality of $G^u$ and $G^d$, the quark mixing matrix $V$ does not become a unit matrix.

	  The role of degenerate symmetry in quark mixing gives a new understanding of the difference between flavor basis and mass basis. 	 
The  charged current weak interaction is introduced in the flavor basis in terms of gauge fields
	\begin{eqnarray}
		\mathcal{L}_{cc}=\frac{g}{\sqrt{2}}\bar{u}_L\gamma_\mu d_L W^+_\mu+h.c.
	\end{eqnarray}
However, we need to diagonalize complex quark mass matrixes to obtain quark mass eigenvalues $q= (U_L)^\dag (G^q)^\dag q^{(m)}$. The Charged current weak interaction becomes
	\begin{eqnarray}
		\mathcal{L}_{cc}=\frac{g}{\sqrt{2}}\bar{u}_L^{(m)} \Big[G^uU_L^u (U_L^d)^\dag (G^d)^\dag\Big]\gamma_\mu d_L^{(m)} W^+_\mu+h.c. 
	\end{eqnarray}
Notice that $G^uU_L^u (U_L^d)^\dag (G^d)^\dag$ is fixed by the CKM measurement. So, the $G^{u,d}$ is not a free transformation, i.e. only a special $G^{u,d}$ is chosen by experiments. 
It means that in the case of mass degeneracy, the CKM matrix is built on a specially chosen mass basis,  not any mass basis. 
Due to this reason, the $G^{q}$ in the degeneracy case is not treated as identity or free parameters. 
This is just the source of misunderstanding on $\delta_{CP}$ vanishing for quark mass degeneracy. 
Generally, $G^{q}$ is a $SU(2)$ transformation between two degenerate mass states. 
 If treating the $SU(2)$ symmetry as a free transformation, CP violation in the CKM matrix can be eliminated by a suitable $SU(2)$ transformation. 
 A detailed calculation on this procession has been given in Appendix \ref{app.SU2}.

The transformation $G^{q}$ can provide CP violation. 
Let us start from a real mixing matrix $U_L^u(U_L^d)^\dag$ and consider $G^q$ role in generating $\delta_{CP}$.

 Generally, a real $U_L^u(U_L^d)^\dag$ is factorized by 3-dimensional orthogonal rotation
	\begin{eqnarray}
	U_L^u(U_L^d)^\dag=\Array{ccc}{1& 0 & 0 \\ 0 & \bar{c}_{23} & \bar{s}_{23} \\ 0 & -s_{23} & \bar{c}_{23}}
		\Array{ccc}{\bar{c}_{13} & 0 & \bar{s}_{13} \\ 0 & 1 & 0 \\ -\bar{s}_{13} & 0 & \bar{c}_{13}}
		\Array{ccc}{\bar{c}_{12} & \bar{s}_{12} & 0 \\ -\bar{s}_{12} & \bar{c}_{12} & 0 \\ 0 & 0 & 1}
	\label{eq.RealUCKM}
	\end{eqnarray}
with $\bar{s}_{ij}=\sin\bar{\theta}_{ij}, \bar{c}_{ij}=\cos\bar{\theta}_{ij}$.
Here, a bar is used just to label mixing angles in the real $U_L^u(U_L^d)^\dag$ to distinguish the mixing angles in phenomenology. 
When $G^{u}=G^{d}=1$, the CKM mixing matrix is completely determined by Eq. (\ref{eq.RealUCKM}), and there is no CP  violation. 

In the presence of a degeneracy in the first two families, $G^q$ can generally be parameterized into a $SU(2)$ transformation
	\begin{eqnarray}
	G^q&=&\Array{ccc}{e^{-i({\phi'}^q+\phi^q)/2}\cos\frac{\theta^q}{2} & -e^{-i({\phi'}^q-\phi^q)/2}\sin\frac{\theta^q}{2} & 0 \\
	e^{i({\phi'}^q-\phi^q)/2}\sin\frac{\theta^q}{2} & e^{i({\phi'}^q+\phi^q)/2}\cos\frac{\theta^q}{2} & 0 \\
	0 & 0 & 1}
	\label{eq.Gu0}
	\end{eqnarray}
with 1 rotation angle $\theta^q$ and 2 phases ${\phi'}^q,\phi^q$.

In terms of Eq. (\ref{eq.UckmGG}),   Jarlskog's invariant can be written as
	\begin{eqnarray*}
	J_{CP}
	&=&\mathcal{C}_1\sin(\phi^d)\sin\theta^d\cos\theta^u
	+\mathcal{C}_2\sin\phi^u\sin\theta^u\cos\theta^d 
	\\
	&&+\mathcal{C}_3\sin(\phi^d+\phi^u)\sin\theta^d\sin\theta^d 
    +\mathcal{C}_4\sin(\phi^d-\phi^u)\sin\theta^d\sin\theta^d 
    	\label{eq.JCPGuGd}
	\end{eqnarray*}
Here, coefficients $\mathcal{C}_i$ are
	\begin{eqnarray*}
	\mathcal{C}_1&=&\frac{1}{2}\bar{c}_{13}^2\bar{s}_{13}\bar{s}_{23}\bar{c}_{23}
		\\
		\mathcal{C}_2&=&\frac{1}{2}\bar{c}_{13}\bar{c}_{23}\Big(
		\bar{s}_{12}\bar{c}_{12}
		+\bar{s}_{12}\bar{c}_{12}\bar{c}_{13}^2\bar{c}_{23}^2
		-2\bar{s}_{12}\bar{c}_{12}\bar{c}_{23}^2
		-2\bar{s}_{23}\bar{c}_{23}\bar{s}_{13}\bar{c}_{12}^2
		+\bar{s}_{23}\bar{c}_{23}\bar{s}_{13}\Big)
		\\
		\mathcal{C}_3&=&\frac{1}{4}\bar{c}_{13} \bar{c}_{23}\Big(\bar{s}_{12}^2\bar{c}_{23}^2-\bar{s}_{12}^2\bar{c}_{13}^2+\bar{c}_{12}^2 \bar{s}_{13}^2\bar{s}_{23}^2 + 
      2\bar{s}_{13}\bar{s}_{12}\bar{c}_{12}\bar{s}_{23}\bar{c}_{23}\Big)
		\\
		\mathcal{C}_4&=&-\frac{1}{4}\bar{c}_{13} \bar{c}_{23} \Big(\bar{c}_{12}^2\bar{c}_{13}^2-\bar{c}_{12}^2\bar{c}_{23}^2-\bar{s}_{12}^2\bar{s}_{13}^2\bar{s}_{23}^2+2\bar{s}_{13}\bar{s}_{12}\bar{c}_{12}\bar{s}_{23}\bar{c}_{23}\Big)
	\end{eqnarray*}
Non-vanishing $J_{CP}$ can be generated by $SU(2)$ transformation from a real orthogonal matrix in Eq. (\ref{eq.RealUCKM}). Phases ${\phi'}^u$ and ${\phi'}^d$ have no contribution to $J_{CP}$ because they can be eliminated by quark field rephasing. 
CP violation requirement to complex phases is provided by $\phi^u$ and $\phi^d$. Another factor affecting CP violation is $SU(2)$ rotation angles $\theta^{u,d}$. A more accurate requirement for CP violation is that at least one of the phases $\phi^{u,d}$ and one of the angles $\theta^{u,d}$ do not vanish, meaning that $G^u$ or $G^d$ can serve as an independent source of CP violation.

Even if $\bar{s}_{13}=0$, there is also non-vanishing $\delta_{CP}$ generated from $G^{u}$ or $G^d$.
For the sake of simplicity, let us take $G^d=1$ and focus on $G^u$. 
The Eq. (\ref{eq.JCPGuGd}) can be simplified to:
	\begin{eqnarray}
		J_{CP}=\frac{1}{2}\bar{c}_{12}\bar{c}_{23}\bar{s}_{12}\bar{s}_{23}^2\sin\phi^u\sin\theta^u
	\end{eqnarray}
So, we get the condition of $\delta_{CP}\neq 0$ is 
	\begin{eqnarray}
		\bar{\theta}_{12}\neq 0, \pi/2 ~\rm{ and}~\bar{\theta}_{23}\neq 0, \pi/2
	\end{eqnarray}
More further result shows that the condition of generating CP violation from real rotation by $G^q$ is that at least two of three mixing angles are not $0$ or $\pi/2$.

Besides the CP violating phase, the contribution of $G^q$ to other mixing angles can be found in Appendix \ref{app.Gqtomixingangles}.

\section{CP violation in  The Minimal Flavor Structure}\label{sec.MFS}
Recently, the minimal flavor structure (MFS) has been proposed to address quark mass hierarchy and CKM mixing in terms of a flat mass pattern \cite{ZhangJPG2023}. 
It successfully outputs 10 experimental values, including 6 quark masses, 3 mixing angles, and 1 CP violation,  from just 10 model parameters.
In the MFS, the CKM matrix is determined by an approximate degenerate symmetry. This example helps us understand the relationship between flavor mixing and mass degeneracy.

As a result of the requirement for mass hierarchy \cite{Zhang2023arXiv3}, the up-type and down-type quark mass matrices in the MFS have a common factorized form as
		\begin{eqnarray}
			M^q&=&m^q_\Sigma Y_L^qI^q(Y_R^q)^\dag
			\label{eq.MFSmass}
		\end{eqnarray}
with the total mass of quarks $m^q_\Sigma=m_1^q+m_2^q+m_3^q$ and diagonal Yukawa matrix $Y_L^q={\rm diag}(e^{i\lambda_1^q} , e^{i\lambda^q_2},1)$. 
A significant aspect of Eq. (\ref{eq.MFSmass}) is that the left-handed $Y_L^q$ fully provides the required complex phases for CP violation.

$I^q$ is responsible for generating hierarchical masses of quarks, which can be represented by a nearly flat real matrix
		\begin{eqnarray}
			I^q&=&I_0^q+\Delta^q
			\nonumber
			\\
			I_0^q&=&\frac{1}{3}\Array{ccc}{1&1&1 \\ 1&1&1 \\ 1&1&1 }
			\nonumber\\
			\Delta^q&=&\frac{1}{3}\Array{ccc}{0 & \delta^q_{12} & \delta^q_{13} \\
			\delta^q_{12} & 0 & \delta^q_{23} \\
			\delta^q_{13 } & \delta^q_{23} &0}
			\label{eq.DeltaDef}
		\end{eqnarray}
In the mass hierarchy limit, quark masses can be gotten by diagonalizing the flat matrix $I_0^q$ by transformation $S_0$
	\begin{eqnarray}
		S_0^q I_0^q (S_0^q)^T={\rm diag}(0,0,1)
		\label{eq.SIS0}
	\end{eqnarray}
with 
	\begin{eqnarray}
		S_0=\frac{1}{\sqrt{6}}\Array{ccc}{\sqrt{3} & 0 & -\sqrt{3} \\ 
			-1 & 2 & -1 \\
			\sqrt{2} & \sqrt{2} & \sqrt{2}} 
	\end{eqnarray}
For 2-fold degenerate eigenvalues on the right side of Equation (\ref{eq.SIS0}), there exists a real $SO(2)$ rotation symmetry $R_0^q$
\begin{eqnarray}
		R^q_0{\rm diag}(0,0,1){R^q_0}^T={\rm diag}(0,0,1)
	\end{eqnarray}
with 
\begin{eqnarray}
R_0^q=\Array{ccc}{\cos\theta^q & \sin\theta^q & 0 \\ -\sin\theta^q & \cos\theta^q & 0 \\ 0 & 0 & 1}
\nonumber
\end{eqnarray}
 Because all complex phases have been factored into $Y_L^q$ in the MFS, the symmetry is only $SO(2)$ rather than $SU(2)$ \cite{XuEPL2023}.
So, we have
	\begin{eqnarray}
		\Big(R_0^qS_0^q\Big)I_0^q\Big(R_0^qS_0^q\Big)^T={\rm diag}(0,0,1)
		\nonumber
	\end{eqnarray}
Thus, the CKM matrix in charged current weak interaction is written in terms of two $SO(2)$ transformations of up-type and down-type quarks as
	\begin{eqnarray}
		V=R_0^uS_0{\rm diag}( e^{i\lambda_1}, e^{i\lambda_2},1)S_0^T{R_0^d}^T 
	\end{eqnarray}

The true hierarchal masses are also addressed by $I^q$ with correction $\Delta^q$ in Eq. (\ref{eq.DeltaDef}) that is responding to generate two lighter quark masses for three real diagonal perturbations $\delta_{ij}^q$. 
Up to $\mathcal{O}(h^1)$, the broken $\delta_{ij}^q$ is set as
	\begin{eqnarray}
	\delta_{12}^q=\delta_{23}^q=-\frac{9}{4}h_{23}^q,~~
	\delta_{13}^q=0
	\end{eqnarray}
$I^q$ is diagonalized by a corrected $S_h^q$
	\begin{eqnarray}
	S_h^qI^q {S_h^q}^T=\Array{ccc}{0&& \\ & h_{23}^q & \\ && 1-h_{23}^q}+\mathcal{O}(h^2) 
	\end{eqnarray}
with  
	\begin{eqnarray}
		S_h^q&=&S_0+\frac{h_{23}^q}{4\sqrt{3}}\Array{ccc}{0 & 0 & 0 \\ \sqrt{2} & \sqrt{2} & \sqrt{2} \\ 1 & -2 & 1}+\mathcal{O}(h^2) 
	\end{eqnarray}

In \cite{Zhang2023arXiv3}, it has been studied that the $SO(2)$ symmetry is still valid
in 1-order hierarchy approximation, i.e., the eigenvalues ${\rm diag}(0, h_{23}^q, 1-h_{23}^q)$ is invariant under an $SO(2)$ transformation $R_h^q$ 
	\begin{eqnarray}
		R_h^q{\rm diag}(0, h_{23}^q, 1-h_{23}^q) {R_h^q}^T={\rm diag}(0, h_{23}^q, 1-h_{23}^q)+\mathcal{O}(h^2)
	\end{eqnarray}
with the transformation 
	\begin{eqnarray}
		R_h^q=R_0^q+\frac{h_{23}^q}{\sqrt{2}}\Array{ccc}{0 & 0 & \sin\theta^q \\ 0 & 0 & \cos\theta^q-1
		\\
		-\sin\theta^q & \cos\theta^q-1 & 0} 
	\end{eqnarray}
So, the CKM mixing matrix in the 1-order hierarchy becomes
	\begin{eqnarray}
		V=R_h^uS_h^u{\rm diag}(e^{i\lambda_1},e^{i\lambda_2},1){S_h^d}^T{R_h^d}^T 
	\end{eqnarray}

The MFS has been successfully checked by fitting hierarchal masses of up-type and down-type quarks and the CKM mixing.  A set of fit parameters is listed in Tab. \ref{tab.fitCKM}.
\begin{table}[htp]
\caption{MFS fit the CKM mixing 
}
\begin{center}
\begin{tabular}{c|c}
\hline\hline
MFS para. & $\theta^u=0.01926$, $\theta^d=3.389$, $\lambda_1=0.004102$, $\lambda_2=-0.04306$
\\
\hline 
Fit Results  & $s_{12}=0.2259$, $s_{23}=0.04172$, $s_{13}=0.003810$, $\delta_{CP}=1.118$
\\
\hline
 CKM exp. \cite{PDG2022} & $\begin{array}{ccc}& s_{12}=0.22500\pm0.00067, & s_{23}=0.04182^{+0.00085}_{-0.00074}
 	\\
	 & s_{13}=0.00369\pm0.00011, &\delta_{CP}=1.144\pm0.027\end{array}$
\\
\hline\hline
\end{tabular}
\end{center}
\label{tab.fitCKM}
\end{table}%

$\theta^u$ and $\theta^d$ listed in Tab. \ref{tab.fitCKM} means that approximate $SO(2)$ symmetry of mass matrix is broken by charged current weak interaction. And $SO(2)$ rotation angles $\theta^u$ and $\theta^d$ are fixed by the SM. 
The $S_h$ transforms the quark flavor basis $q_L$  to a mass basis $q_L^{(m)}$
$$q_L=S_h^T q_L^{(m)}$$
And charged current weak interaction decides which mass basis appears in the CKM mixing. So, only the fitted $\theta^{u,d}$ is chozen
\begin{eqnarray}
	u_L&=&S_h^T{R_h^u}^T u_L^{(m)},
	\\
	d_L&=&S_h^T{R_h^d}^T d_L^{(m)}
\end{eqnarray}

MFS also provides an understanding of the origin of CP violation. CP violation comes from Yukawa phases in $Y_L^q$ and also depends on rotation angles $\theta^u,\theta^d$. These parameters are independent on mass hierarchy, They do not vanish in the mass hierarchy limit, which explains why the large CP violating phase does not stem from the mass hierarchy.

\section{Summary}
\label{sec.summ}
The study of degenerate symmetry as an approximation of mass hierarchy has been studied in quark CKM mixing, particularly in CP violation. Degenerate symmetry $G^q$ plays a non-trivial role in CKM mixing and picks up a special mass eigenstate as the state of the CKM mixing matrix in charged current weak interaction. 
It explains why CP violation may not vanish for mass degeneracy and also why a large CPV can exist in the hierarchy limit. As an example, the role of degenerate symmetry is shown in the MFS. The relation between CP violation and degenerate symmetry is also illustrated in the MFS. These results assist in improving the understanding of CP violation and aid in constructing a final flavor structure in the future.

\section*{Acknowledgements}
This work is supported by Shaanxi Foundation SAFS 22JSY035 of China.


\begin{appendix}
\section{Elimination of  $\delta_{CP}$ by a free $SU(2)$ transformation}
\label{app.SU2}	
Phenomenology, the CKM mixing matrix $V$ can be expressed by 3 mixing angles and 1 CP violating phase by redefining quark fields. This process is known as rephasing.  For degenerate masses, there is a $SU(2)$ symmetry to keep mass eigenvalues invariant. As we have mentioned in Sec. \ref{sec.degeneratesym}, these $SU(2)$ transformations, $G^u$ and $G^d$, are decided by experiments. However, if $G^{u,d}$ is regarded as free transformation, it can be utilized to eliminate some d.o.f. in the CKM mixing. In the appendix, we adopt the standard CKM matrix in Eq. (\ref{eq.DefCKM}) and demonstrate the elimination of $\delta_{CP}$ through a free $SU(2)$ transformation.
Without the loss of generality, we only consider up-type quark degeneracy between the first two families. Degenerate symmetry $G^u$ is parameterized by $\theta^u$, $\phi^u$ and $\phi'^u$ as shown in Eq. \ref{eq.Gu0}.

To obtain a vanishing CP violation, a transformed CKM matrix $G^{u}V$ must have a vanishing Jarlskog invariant. After a tedious calculation, $J_{CP}$ is expressed by
	\begin{eqnarray}
	J_{CP}=\mathcal{C}_a\cos\theta^u + \mathcal{C}_b\sin\phi^u\sin\theta^u + \mathcal{C}_c\cos\phi^u\sin\theta^u
	\end{eqnarray}
with coefficients
	\begin{eqnarray*}
		\mathcal{C}_a&=&c_{13}^2s_{13}c_{23}s_{23} c_{12}s_{12} s_\delta
		\\
		\mathcal{C}_b
		&=&\frac{1}{2}c_{13}c_{23} 
			\Big[s_{12}c_{12}(c_{13}^2-c_{23}^2 
				+s_{13}^2s_{23}^2-2s_\delta^2s_{13}^2s_{23}^2) 
			- c_\delta(2c_{12}^2-1) s_{13}s_{23}c_{23}\Big]
		\\
		\mathcal{C}_c
		&=&\frac{1}{2}c_{13}c_{23}  s_{13} s_\delta\Big[2s_{12}c_{12}s_{13}s_{23}^2c_\delta- s_{23}c_{23}(2c_{12}^2-1)\Big]
     	\end{eqnarray*}
Here, mixing angles $\theta_{ij}$ and $\delta_{CP}$ are ones before $SU(2)$ transformation.

Vanishing $J_{CP}$ requires $\theta^u$ and $\phi^u$ meet
	\begin{eqnarray}
		\tan\theta^u=-\frac{\mathcal{C}_a}{\mathcal{C}_b \sin\phi^u +\mathcal{C}_c \cos\phi^u}
	\end{eqnarray}

\section{ $G^q$ roles to mixing angles}
\label{app.Gqtomixingangles}
Eq. (\ref{eq.JCPGuGd}) has expressed degenerate symmetry $G^q$ contribution to Jarlskog invariant initializing from a real mixing matrix.
We will now examine the influence of $G^q$ on the remaining three mixing angles.
Three CKM mixing angles can be calculated from mixing matrix $V$ 
	\begin{eqnarray}
		s_{13}&=&|V_{13}|
		\\
		s_{12}^2&=&\frac{|V_{12}|^2}{1-|V_{13}|^2}
		\\
		s_{23}^2
&=&\frac{|V_{23}|^2}{1-|V_{13}|^2}
	\end{eqnarray}
Let us consider a real $U_L^u(U_L^d)^\dag$ as shown in Eq. (\ref{eq.RealUCKM}). Using Eq. (\ref{eq.UckmGG}) and Eq. (\ref{eq.Gu0}), three mixing angles are determined by
	\begin{eqnarray*}
	s_{13}^2&=&\frac{1}{2}(1+\bar{c}_u)\bar{s}_{13}^2 + \bar{c}_{13}\bar{s}_{23}\Big[\bar{c}_{13}\bar{s}_{23}\frac{1}{2}(1-c_u) - \bar{s}_{13}s_u\cos\phi^u\Big]
		\\
	s_{12}^2(1-s_{13}^2)
	&=&\mathcal{C}_{A1}+ \mathcal{C}_{A2} c_d + \mathcal{C}_{A3} s_d +\mathcal{C}_{A4} c_u + \mathcal{C}_{A5} c_u c_d + \mathcal{C}_{A6} c_u s_d 
	\\
	&&+
\mathcal{C}_{A7} s_u + \mathcal{C}_{A8} s_u c_d + \mathcal{C}_{A9} s_u s_d
		\\
	s_{23}^2(1-s_{13}^2)
	&=&\frac{1}{2} \Big[ \bar{s}_{13}^2 
		+\bar{c}_{13}^2 \bar{s}_{23}^2
		+ (\bar{c}_{13}^2 \bar{s}_{23}^2--\bar{s}_{13}^2 )c_u
 	 +2\bar{s}_{13}\bar{c}_{13}  \bar{s}_{23}s_u\cos\phi^u\Big]
	\end{eqnarray*}
Here, coefficients $\mathcal{C}_{Ai}$ are listed in the following
	\begin{eqnarray*}
		\mathcal{C}_{A1}
			&=&\frac{1}{4}(2 - \bar{s}_{23}^2 - \bar{s}_{13}^2 \bar{c}_{23}^2)
		\\
		\mathcal{C}_{A2}&=&-\bar{c}_{12} \bar{c}_{23} \bar{s}_{12} \bar{s}_{13} \bar{s}_{23} +\frac{1}{4} (1- 2 \bar{s}_{12}^2) (\bar{s}_{13}^2-\bar{s}_{23}^2- \bar{s}_{13}^2 \bar{s}_{23}^2)
		\\
		\mathcal{C}_{A3}&=&\frac{1}{2} \Big[\bar{c}_{23} (-1 + 2 \bar{s}_{12}^2) \bar{s}_{13} \bar{s}_{23} + 
   \bar{c}_{12} \bar{s}_{12} (\bar{s}_{23}^2 -\bar{s}_{13}^2 +\bar{s}_{13}^2\bar{s}_{23}^2)\Big] \cos\phi^d
  		 \\
		\mathcal{C}_{A4}&=&\frac{1}{4} (\bar{s}_{23}^2 - \bar{s}_{13}^2 - \bar{s}_{13}^2\bar{s}_{23}^2)
  		 \\
		\mathcal{C}_{A5}&=&\bar{c}_{12} \bar{c}_{23} \bar{s}_{12} \bar{s}_{13} \bar{s}_{23} - 
 \frac{1}{4}(-1 + 2 \bar{s}_{12}^2) (-2 + \bar{s}_{23}^2 + \bar{s}_{13}^2 +\bar{s}_{13}^2 \bar{s}_{23}^2)
  		 \\
		\mathcal{C}_{A6}&=&\frac{1}{2} \Big[\bar{c}_{23} (1- 2 \bar{s}_{12}^2) \bar{s}_{13} \bar{s}_{23} + 
   \bar{c}_{12} \bar{s}_{12} (2 -\bar{s}_{23}^2 -\bar{s}_{13}^2 -\bar{s}_{13}^2 \bar{s}_{23}^2)\Big] \cos\phi^d
  		 \\
		\mathcal{C}_{A7}&=&\frac{1}{2} \bar{c}_{13} \bar{s}_{13} \bar{s}_{23} \cos\phi^u
  		 \\
		\mathcal{C}_{A8}&=&-\frac{1}{2} \bar{c}_{13} \Big[2 \bar{c}_{12} \bar{c}_{23} \bar{s}_{12} + (1 - 2 \bar{s}_{12}^2) \bar{s}_{13} \bar{s}_{23}\Big]\cos\phi^u
  		 \\
		\mathcal{C}_{A9}&=& \frac{1}{2}\bar{c}_{13}(2\bar{c}_{23} \bar{s}_{12}^2-\bar{c}_{23}  + 2 \bar{c}_{12} \bar{s}_{12} \bar{s}_{13} \bar{s}_{23}) \cos\phi^d\cos\phi^u- \frac{1}{2}\bar{c}_{13}\bar{c}_{23} \sin\phi^d\sin\phi^u
	\end{eqnarray*}
\end{appendix}



\begin{thebibliography}{99}
\bibitem{Hocker2006}
	A. Hocker, Z. Ligeti, Ann. Rev. Nucl. Part. Sci. 56 (2006) 501-567 [arXiv: hep-ph/0605217].
\bibitem{Isidori2010}
	G. Isidori, PoS ICHEP2010 (2010) 543 [arXiv: 1012.1981].
\bibitem{Zupan2019}
	J. Zupan, CERN Yellow Rep.School Proc. 6 (2019) 181-212 [arXiv: 1903.05062].
\bibitem{Raidal2008}
	M. Raidal, et al, Eur.Phys.J.C 57 (2008) 13-182 [arXiv:  0801.1826].
\bibitem{Jarlskog1985}
	C. Jarlskog, Phys. Rew. Lett. 55 (1985) 10.
\bibitem{PDG2022}
	R.L. Workman et al. (Particle Data Group), Prog. Theor. Exp. Phys. 2022, 083C01 (2022) and 2023 update.
\bibitem{ZhangJPG2023}
	Y. Zhang, J.Phys.G 50 (2023) 12, 125006 [arXiv: 2302.05943].
\bibitem{Zhang2023arXiv3}
	Y. Zhang, arXiv:2311.09044.
\bibitem{XuEPL2023}
	G. Xu, Y. Zhang, EPL 143 (2023) 4, 44001[arXiv: 2304.05017].
\end{thebibliography}
\end{document}